\documentclass[aps,prd,twocolumn,showpacs,superscriptaddress,groupedaddress]{revtex4-2}  
\usepackage{graphicx}  
\usepackage{dcolumn}   
\usepackage{bm}        
\usepackage{amssymb}   
\usepackage{amsmath}   

\usepackage{color} 

\usepackage{multirow}
\usepackage{enumerate}
\usepackage{comment}
\usepackage{tabularx}
\usepackage{booktabs}
\graphicspath{{./figures/}}

\usepackage{hyperref}
\hypersetup{
	colorlinks=true, 	
	linkcolor=cyan,		
    citecolor=red,		
}

\newcommand{\irm}{{\rm i}}

\newcommand{\FLAG}[1]{{\color{red} #1}}

\def \myTitle {Characterization of the seismic field at Virgo and improved estimates of Newtonian-noise suppression by recesses}

\begin{document}

\title{\myTitle}

\author{Ayatri~Singha}
\email{a.singha@maastrichtuniversity.nl}
\affiliation{Department of Gravitational Waves and Fundamental Physics, Maastricht University, P.O. Box 616, 6200 MD Maastricht}
\affiliation{Nikhef, Science Park 105, 1098 XG Amsterdam, The Netherlands}

\author{Jan~Harms}
\affiliation{Gran Sasso Science Institute (GSSI), I-67100 L'Aquila, Italy}
\affiliation{INFN, Laboratori Nazionali del Gran Sasso, I-67100 Assergi, Italy}
\author{Stefan~Hild}
\affiliation{Department of Gravitational Waves and Fundamental Physics, Maastricht University, P.O. Box 616, 6200 MD Maastricht}
\affiliation{Nikhef, Science Park 105, 1098 XG Amsterdam, The Netherlands}

\author{Maria~C.~Tringali}%
\affiliation{European Gravitational Observatory (EGO),
I-56021 Cascina, Pisa, Italy}
\author{Federico~Paoletti}%
\affiliation{INFN, Sezione di Pisa, I-56127 Pisa, Italy}
\author{Irene~Fiori}%
\affiliation{European Gravitational Observatory (EGO), I-56021 Cascina, Pisa, Italy}
\author{Tomasz Bulik}
\affiliation{Astronomical Observatory University of Warsaw, 00-478 Warsaw, Poland}
\affiliation{Astrocent, Nicolaus Copernicus Astronomical Center, Rektorska 4, 00-614 Warsaw Poland}
\author{Bartosz Idzkowski}
\affiliation{Astronomical Observatory University of Warsaw, 00-478 Warsaw, Poland}
\author{Alessandro Bertolini}
\affiliation{Nikhef, Science Park, 1098 XG Amsterdam, Netherlands}

\author{Enrico Calloni}
\address{Universit\`a di Napoli “Federico II”, Complesso Universitario di Monte S.Angelo, I-80126 Napoli, Italy}
\address{INFN, Sezione di Napoli, Complesso Universitario di Monte S.Angelo, I-80126 Napoli, Italy}
\author{Luciano Errico}
\address{INFN, Sezione di Napoli, Complesso Universitario di Monte S.Angelo, I-80126 Napoli, Italy}
\author{Rosario De Rosa}
\address{Universit\`a di Napoli “Federico II”, Complesso Universitario di Monte S.Angelo, I-80126 Napoli, Italy}
\address{INFN, Sezione di Napoli, Complesso Universitario di Monte S.Angelo, I-80126 Napoli, Italy}

\author{Alberto Gennai}
\address{INFN, Sezione di Pisa, Edificio C -- Polo Fibonacci -- Largo B.~Pontecorvo 3, I-56127 Pisa, Italy}



\begin{abstract}
Fluctuations of gravitational forces cause so-called Newtonian noise (NN) in gravitational-wave (GW) detectors which is expected to limit their low-frequency sensitivity in upcoming observing runs. Seismic NN is produced by seismic waves passing near a detector's suspended test masses. It is predicted to be the strongest contribution to NN. Modeling this contribution accurately is a major challenge. Arrays of seismometers were deployed at the Virgo site to characterize the seismic field near the four test masses. In this paper, we present results of a spectral analysis of the array data from one of Virgo's end buildings to identify dominant modes of the seismic field. Some of the modes can be associated with known seismic sources. Analyzing the modes over a range of frequencies, we provide a dispersion curve of Rayleigh waves. We find that the Rayleigh speed in the NN frequency band 10\,Hz--20\,Hz is very low ($\lesssim$100\,m/s), which has important consequences for Virgo's seismic NN. Using the new speed estimate, we find that the recess formed under the suspended test masses by a basement level at the end buildings leads to a 10 fold reduction of seismic NN. 

\end{abstract}

\maketitle

\section{Introduction}

The interferometric gravitational-wave (GW) detectors LIGO and Virgo have brought exciting discoveries in the past years \cite{AbEA2016a,AbEA2017d,MM170817,PhysRevX903104, VIRGO_interferometry, aLIGO_harware, aLIGO_seismic_isolation}. Newtonian noise (NN) is one of the noises predicted to limit the sensitivity of GW detectors in the future at frequencies below 20\,Hz \cite{Sau1984,Har2019,AmEA2020}. Low-frequency sensitivity is important for observations of black-hole and neutron-star binaries alike events\cite{LyEA2015,ChEA2018,HaHi2018}. Terrestrial NN arises from the density perturbation of the environment associated for example with seismic waves \cite{Har2019}. These density fluctuations produce a fluctuating gravitational force that randomly pushes the suspended test masses of a GW detector. Techniques have been proposed to reduce NN \cite{HuTh1998,Cel2000,HaHi2014}. 

One way to reduce NN is by going underground where NN contributions from surface Rayleigh waves and the atmosphere are smaller \cite{BeEA2010,BaHa2019}. This is why Einstein Telescope, a new infrastructure for future GW detectors, is proposed to be underground \cite{Homstake_work1, ET2011, Homstake_work}. Alternatively, one can deploy sensor arrays to use their data for a coherent subtraction of NN from GW data \cite{Cel2000,HaVe2016,CoEA2018a}. The efficiency of NN cancellation (NNC) depends on the number of seismometers, what type of sensor, their sensitivity, their placements around the test masses, the non-stationary behavior of noise sources, etc \cite{CoEA2016a,LIGO_HanObs2020,BaEA2020}. A NNC system is under development for the Virgo detector, which required a detailed characterization of seismic fields at the site \cite{TrEA2019,BaEA2020}.

\begin{figure*}[htbp!] 
    \centering
    \includegraphics[width=\textwidth]{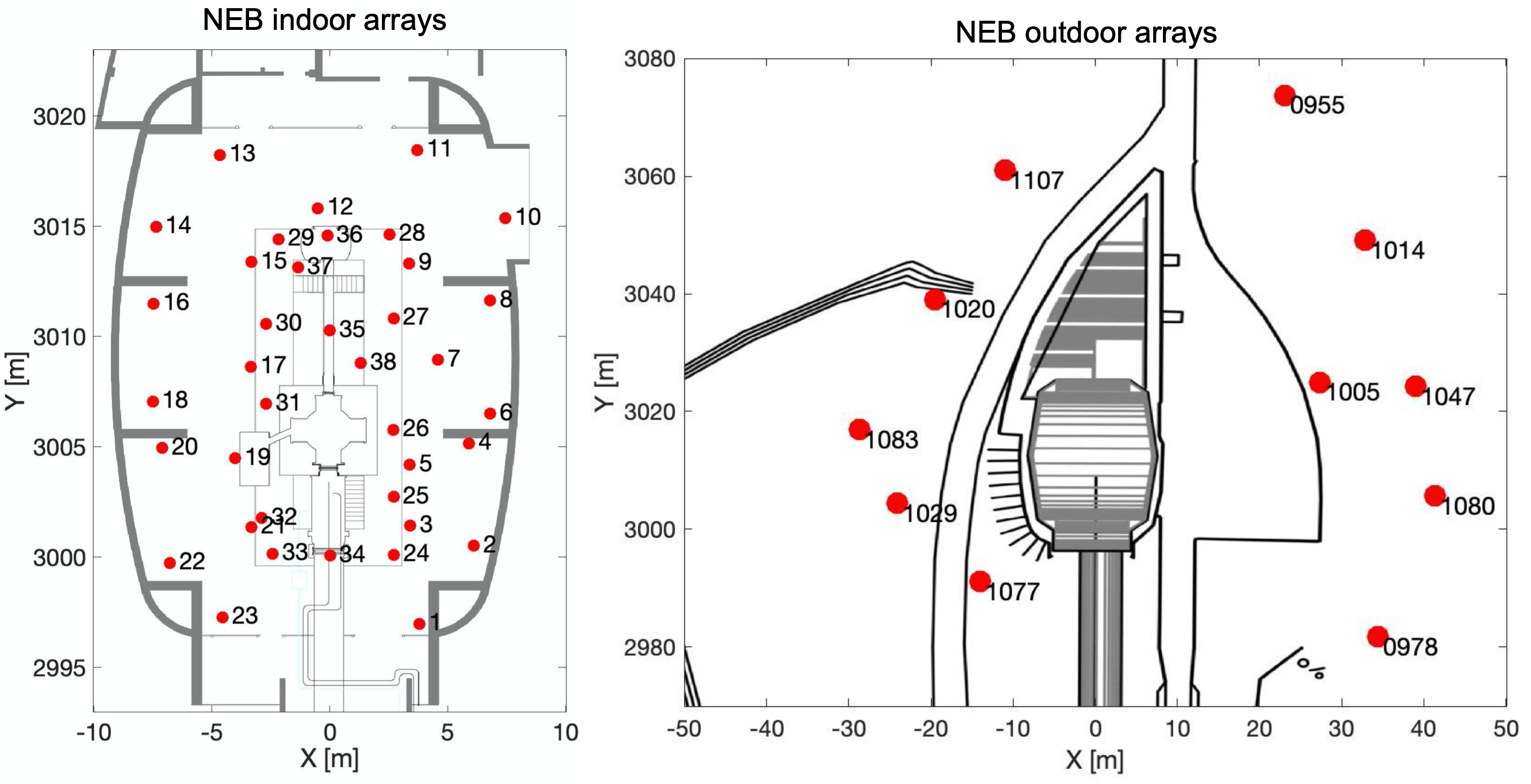}
\caption{Positions of seismometers at the North End building (NEB). (Left) indoor array. (Right) outdoor array. }
\label{fig:Array_pos}
\end{figure*}

In this paper, we present a spectral analysis (in space and time) of the seismic field at the North End Building (NEB) of the Virgo detector. Data used for this analysis were obtained from an array of 38 indoor sensors and 11 outdoor sensors as shown in figure \ref{fig:Array_pos}. The spectral analysis allows us to determine the propagation direction and speed of seismic waves over some frequency band. Sources of dominant contributions to the field are thereby identified. We obtain evidence for scattering of seismic waves, and measure the dispersion curve of Rayleigh waves. The latter can be compared with previous results of array measurements carried out outside the buildings between the two interferometer arms  \cite{sVirgo2017}.

Apart from the NNC, we can reduce NN by modifying local geometry around the detector test mass. In 2014, Harms and Hild showed that if we can build a trench-like structure around a test mass, we can have a significant reduction in NN simply because there is less dense material to support ground vibrations \cite{HaHi2014}. Also, they have discussed the parameters of the trench and pillar system to optimize the NN reduction. In Virgo, clean rooms under the input and end test masses were built for installation of the suspended payloads. It was shown that there should be a reduction in NN by at least a factor 2 between 10\,Hz and 20\,Hz just from the presence of these rooms \cite{VNNReasses2020}. However, certain assumptions had to be made for the analyses. Most importantly, the seismic speed is a critical parameter for the effect of a recess, which was assumed to be 250\,m/s adopting the speed observed at the LIGO sites. In our new analysis presented here, we recalculate the recess NN reduction for Virgo based on the observed Rayleigh-wave dispersion.

This paper is organized as follows. In section \ref{SSRCharacterisation}, we describe the spatial spectra (also referred to as kf maps) produced for different frequencies from the spectral correlation of the seismometer array in the Virgo NEB building. We explicitly discuss such kf maps for three frequencies 10\,Hz, 15\,Hz, and 20\,Hz. In section \ref{Virgo_NN_reassessment}, we first show the velocity dispersion curve based on the dominant noise components in the kf maps generated at different frequencies. Using the estimated dispersion curve, we reassess the NN estimate for Virgo. In section \ref{Conclusion}, we discuss and conclude our results.

\section{Characterisation of the seismic field}
\label{SSRCharacterisation}
In this section, we present an array analysis in wave-vector space ($\vec k=(k_x,\,k_y)$) at different temporal frequencies, which can be used to identify potential local seismic sources. We consider the data from the Virgo NEB seismometers. We first estimate the cross-spectral density $C(\vec r_i, \vec r_j, \omega)$ between the pairs of seismometers. Here, $\vec r_i$ is the positions of seismometer $i$. Hence, for a particular frequency, $C(\vec r_i, \vec r_j)$ spans an $N\times N$ matrix of cross-correlations between $N$ seismometers. We then estimate a spatial spectrum from this matrix \cite{KrVi1996,LIGO_HanObs2020}:
\begin{equation}
    p(\omega, \vec{k}) = \sum_{i, j = 1}^{N}C(\omega; \vec r_i, \vec r_j)\,e^{-\irm\vec k\cdot(\vec r_i - \vec r_j)}
\end{equation}

\begin{figure}[ht!]
    \includegraphics[width=0.5\textwidth]{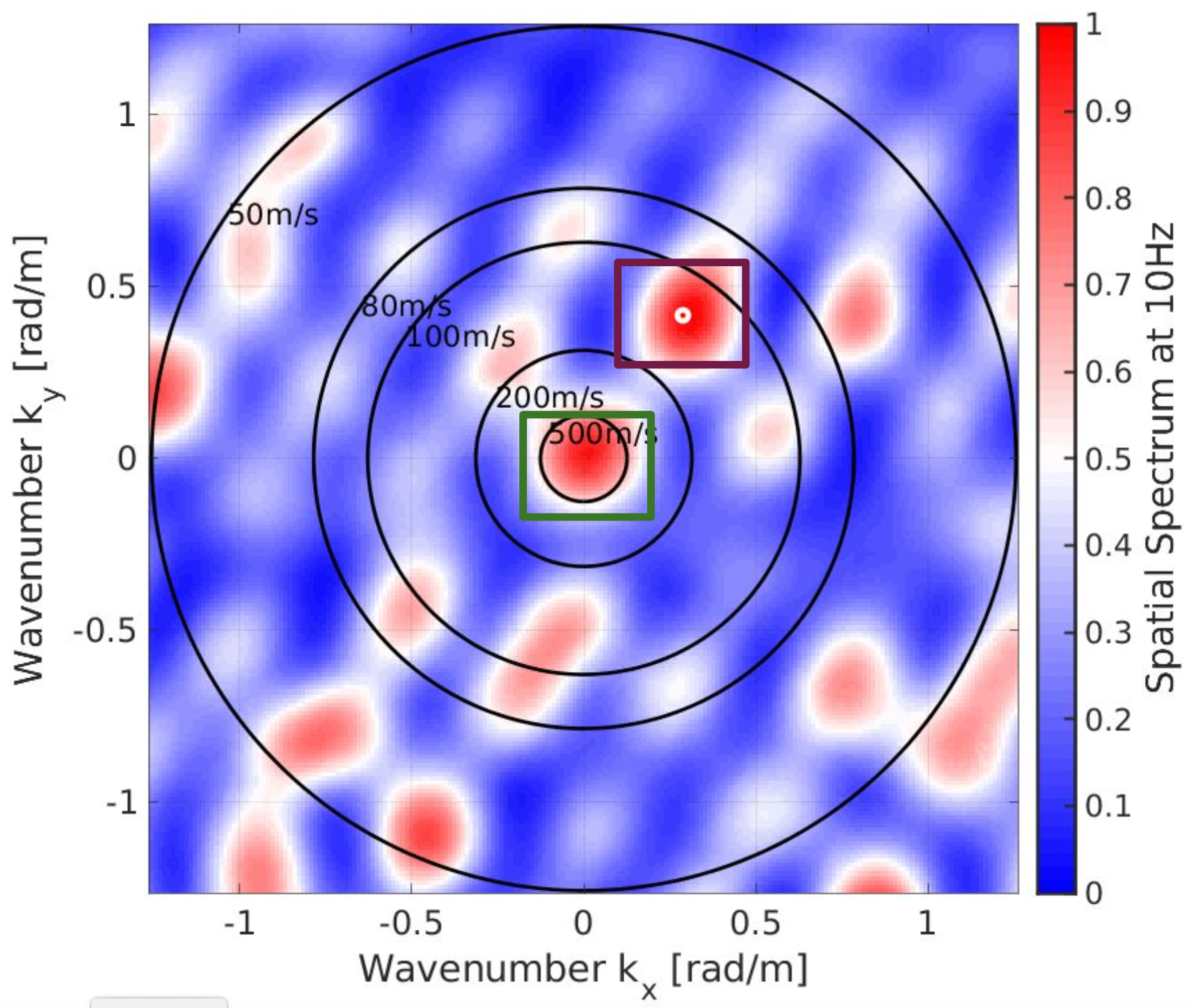}
    \caption{Spatial spectrum at 10\,Hz obtained from cross-correlations between the seismometers and normalized by the maximum value. }
    \label{fig:SS10Hz}
\end{figure}

The matrix C is Hermitian, and hence $p$ is real-valued. For our analysis, we calculate such a spectrum for each hour. We have considered two weeks of data. The spatial spectrum $p$ is the result of an average over many short-term cross-spectral densities $C$. The spatial spectrum $p(\vec k)$ can be calculated for any frequency, but there is a useful range of spatial frequencies connected to the density of the array and its diameter. With some seismometers being as close as 1\,m to each other, and with an array diameter of about 20\,m along the X direction, wavenumbers up to 3\,rad/m can be analyzed with a resolution of about 0.15\,rad/m. This can provide us information about the dominant modes in the spectra and their directionality. A first example of a spatial spectrum is shown in figure \ref{fig:SS10Hz} for a frequency of 10\,Hz. The spectrum is normalized by its maximum value. We see there are several modes (as red blob) present simultaneously. The circles in the plot represent a set of phase speeds, which allows us to estimate by inspection the speed of the modes. 

At 10\,Hz, we observe two strong, distinct modes (boxed) which propagate along the same direction (not clearly visible in the plot). One mode propagates at a speed of approximately 100\,m/s and can therefore be identified as Rayleigh wave. The other mode has a phase speed of a few km/s, which must be a body wave. As both the body wave and the Rayleigh wave come from similar directions, one can suspect that these two modes are generated by the same seismic source.

\begin{figure}[ht!]
    \includegraphics[width=0.5\textwidth]{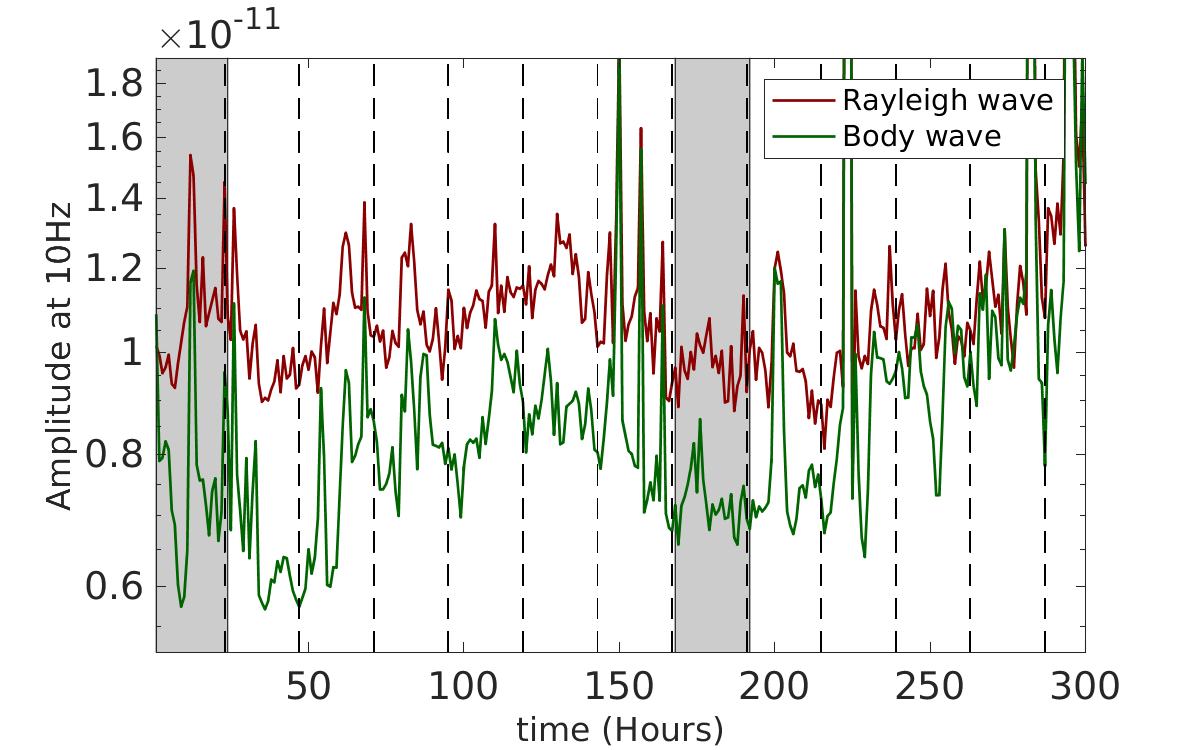}
  \caption{Amplitude variation of dominant 10\,Hz Rayleigh-wave and body-wave modes w.r.t. time (in hours). Black dotted lines represent midnight in local time. The amplitude of the Rayleigh waves (maroon curve) and the body waves (green curve) are correlated and follow a day-night cycle. The waves tend to arrive less frequently at weekends, here around hours 50 and 200 (grey patch), leading to lower amplitudes in the 10\,Hz spectra.
 }
  \label{fig:ampRWBW}
\end{figure}
We have also studied the amplitude variation of these two modes over time. For this, we produce the spatial spectrum $p(\vec k)$ at 10\,Hz averaging over one hour of data and repeat this for a total of 15 days. We find that the two waves are present continuously, and we thus collect the amplitude of the Rayleigh and body waves for each hour. We show the amplitude variation of the Rayleigh and body waves in figure \ref{fig:ampRWBW}. We observe that the amplitude variation of the two modes is similar. We should not expect a perfect correlation between amplitudes of Rayleigh and body waves even if they come from the same source as their propagation is different. Some difference can also be due to other sources occasionally producing stronger Rayleigh and body waves at 10\,Hz. The magnitude of the body wave is always lower than the Rayleigh wave, which is expected for surface sources. The dotted lines in the figure \ref{fig:ampRWBW} represent midnight in local time. Strong disturbances only happen during the day. The location of a highway bridge with respect to the Virgo NEB matches with the direction of observed Rayleigh and body waves. Previous studies of the bridge noise focused on lower frequencies (below 5\,Hz) and on the analysis of spectrograms where the two waves appear as frequent transients \cite{FHP2003,KoEA2017}. Assuming that these waves come from the highway bridge, we can expect this noise to be high during the day when traffic is heavier on the highway. 
\begin{figure}[ht!]
    \includegraphics[width=0.5\textwidth]{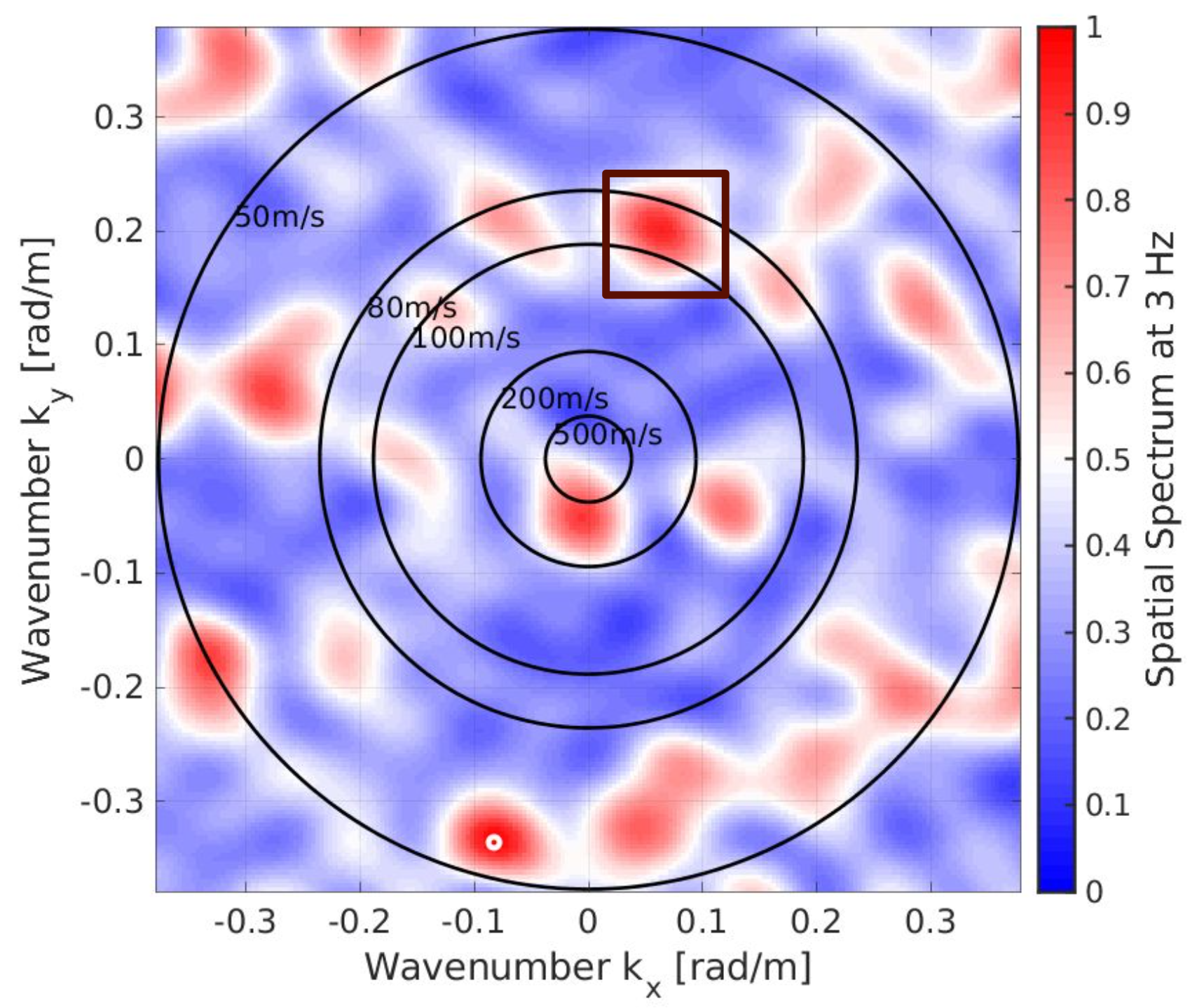}
  \caption{Spatial spectrum at 3\,Hz drawn from cross-correlations between the 11 outdoor seismometers normalized by the maximum value. The boxed region represents the Rayleigh-wave mode which we have observed already at 10\,Hz using the indoor sensors. }
  \label{fig:ExtSens3Hz}
\end{figure}

The position of the indoor and outdoor arrays is shown in figure \ref{fig:Array_pos}. In figure \ref{fig:ExtSens3Hz} we show the 3\,Hz spatial spectrum for the outdoor array. We again see a Rayleigh wave coming from the same direction as the highway bridge. Thus, as already concluded in past studies, there is strong evidence that the highway bridge is an important noise source in the Virgo environment.

\begin{figure*}[htbp!] 
    \centering
    \includegraphics[width=\textwidth]{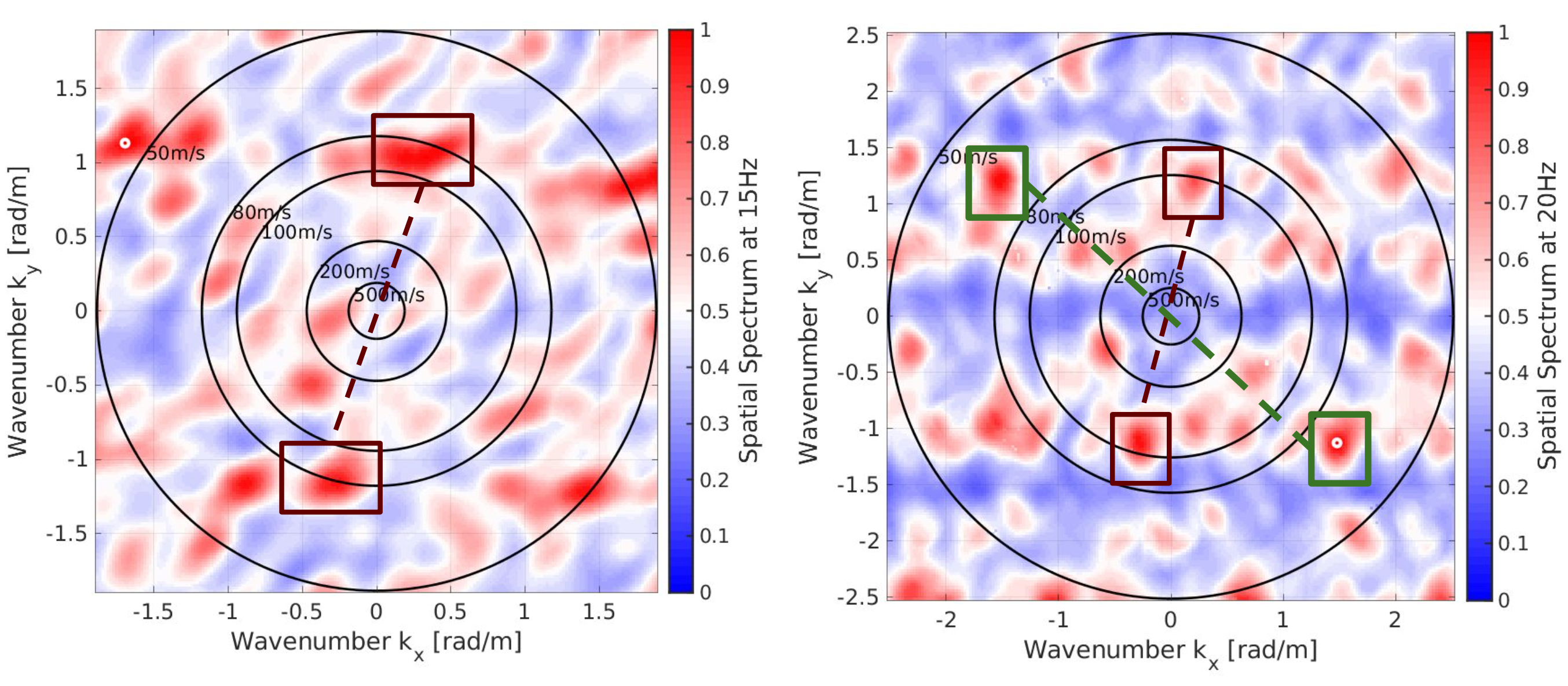}
\caption{Spatial spectra at 15\,Hz (left) and 20\,Hz (right) normalized by their maximum values. The boxes mark counter propagating twins, where one is an artifact produced by aliasing.}
\label{fig:SS_15_20Hz}
\end{figure*}

\begin{figure}[ht!]
    \centering
    \includegraphics[width=9cm,height=8cm,keepaspectratio]{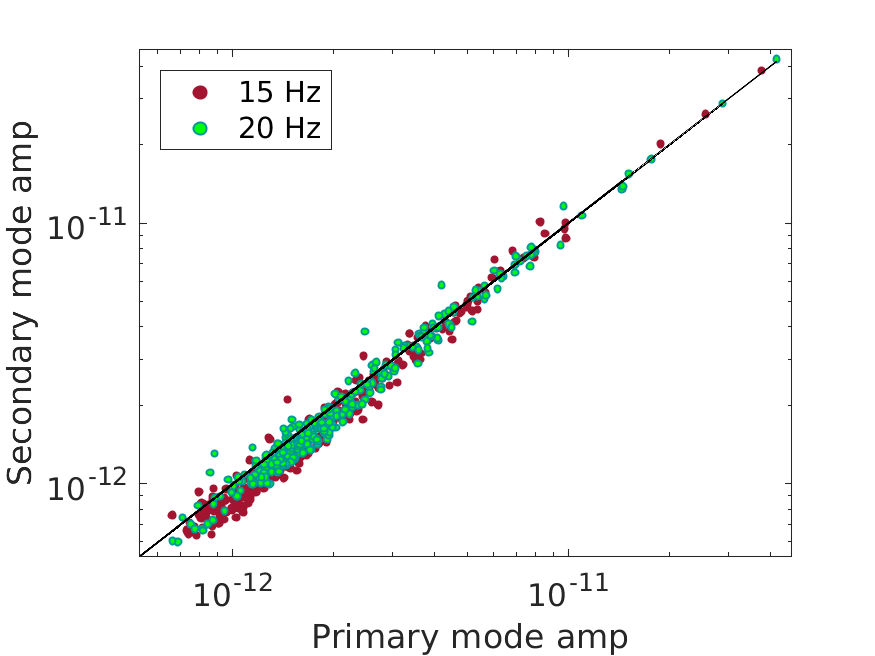}
\caption{Amplitudes of primary and secondary modes shown in figure \ref{fig:SS_15_20Hz}. Maroon dotted points represent the amplitude of counter-propagating twins observed at 15\,Hz and green dotted points represent the amplitudes of the counter-propagating twins boxed in figure \ref{fig:SS_15_20Hz}  at 20\,Hz. The black solid line is the straight line of equality (y = x). }
\label{fig:Amp_15_20Hz}
\end{figure}

In figure \ref{fig:SS_15_20Hz}, we show similar plots for 15 and 20\,Hz. Here, the spectra are more complicated since aliasing of modes starts to play a role. Aliasing is a phenomenon where fake modes appear in the spectrum due to finite spatial resolution and bandwidth of the array. At 15\,Hz a counter-propagating twin of waves appears (boxed modes in the left plot). One might conclude that one wave could be the reflection of the other or two seismic sources at opposite directions produce waves simultaneously, but in fact, the properties of the twin are consistent with an aliasing effect. As explained earlier, the largest analyzable wavenumber is about 3\,rad/m, which is approximately the distance between the twin modes. At 20\,Hz shown in the right plot, the aliasing pattern is clearly visible, i.e., identifiable as artificial pattern in the spatial spectrum. Two twin modes are marked in boxes, and again, the distance between them corresponds roughly to the spectral bandwidth of the array analysis (given by the sensor density). 

So, it turns out that Rayleigh waves are so slow at the Virgo site that it is not possible to carry out a good spatial spectral analysis well above 10\,Hz with the NEB array. Its sensors would have to be moved closer together, but then it would be more difficult to analyze the fast body waves. The best option would be to add more sensors so to increase the sensor density while maintaining the array diameter (and therefore the spatial resolution). For some analyses though, e.g., to measure the phase speed of a mode, it is possible to simply select one of the two modes of a twin for the analyses since the real wave and its twin alias in the Virgo spatial spectra have the same speeds. However, since the array configuration is irregular, the amplitudes of the aliases are only similar but not the same as of the real waves. Since the dominant modes at 15\,Hz and 20\,Hz are persistent, we can analyze them over a longer period of time. The amplitudes of the two modes of a twin are confronted in figure \ref{fig:Amp_15_20Hz}. The plot shows that the amplitudes of these twin modes are of  same order and strongly correlated as they are scattered around the line of equality. Here, their relative amplitudes can be smaller than 1 if the modes picked for the analysis are not the dominant contribution to the respective spatial spectra.

\begin{figure}[ht!]
    \centering
    \includegraphics[width=0.95\columnwidth]{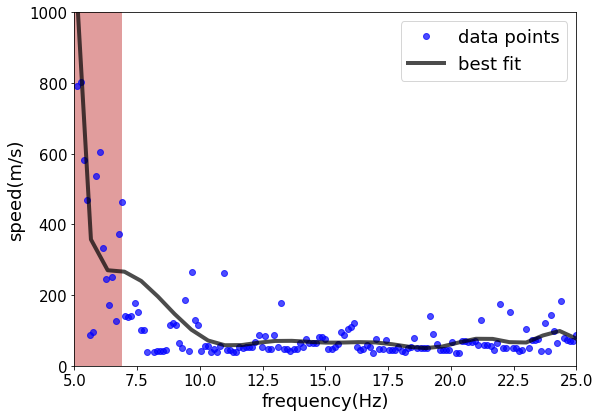}
  \caption{Rayleigh dispersion plot showing average speed of versus frequency. Blue dots represent the data points, the black curve is a fit to the data. The red shaded region marks frequencies outside the array bandwidth (seismic waves are too long to be analyzed). }
  \label{fig:VelsDisp}
\end{figure}

A tiltmeter was present at the NEB together with the seismometer array, which allows us to do an additional analysis important to NNC. The tiltmeter is predicted to play a special role for the cancellation of NN from Rayleigh waves \cite{HaVe2016}. If a field is composed entirely of plane Rayleigh waves, then the associated NN can be canceled by a single tiltmeter instead of using an array of seismometers. A first check to find out whether a field can be considered (approximately) as composed by plane Rayleigh waves is to compare the array-inferred ground tilt with the ground tilt measured by a tiltmeter. Limitations of array analyses (aliasing and resolution limits) make the comparison difficult, but a good match between the two was observed at the LIGO Hanford site \cite{LIGO_HanObs2020}, which is a strong indication that a tiltmeter would be very effective for NNC at Hanford.

\begin{figure}[ht!]
    \centering
    \includegraphics[width=0.95\columnwidth]{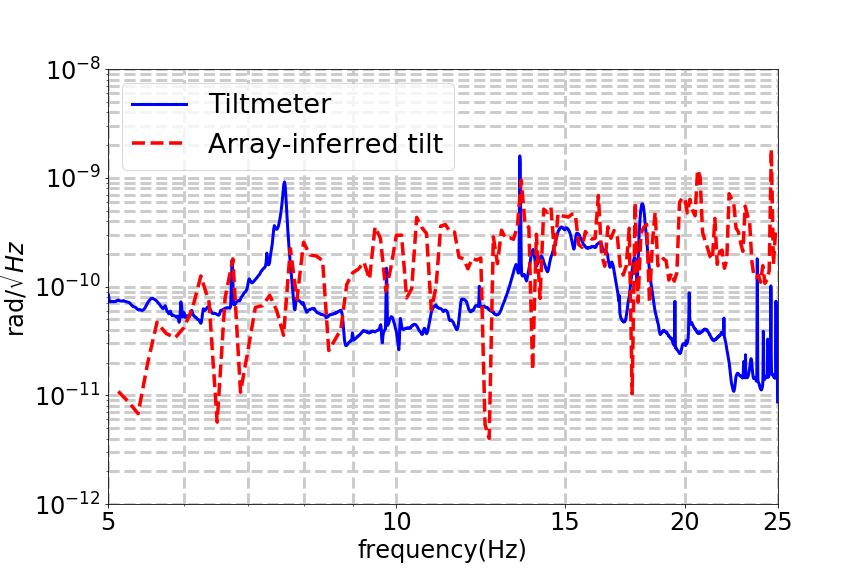}
  \caption{The plot shows the estimated ground tilt spectra (red dotted trace) and the spectra directly estimated from ground tiltmeter data (blue solid trace). }
  \label{fig:TiltSpectra}
\end{figure}

There are various approaches and levels of approximation to estimate the ground tilt from array data. We base our analysis of the dominant modes in the spatial spectrum, estimate their wavenumber $k$ and propagation direction $\phi$, and then average over many estimates. The power spectral density of the array-inferred ground tilt can then be written as
  \begin{equation}
      S(\tau_x;\omega) = <k^2 \cos^2(\phi)> S(\xi;\omega),
  \end{equation}
where $S(\xi;\omega)$ is the spectral density of vertical ground displacement. The result is shown in figure \ref{fig:TiltSpectra}. The overall level is similar, and one might conclude the the match around 15\,Hz is good indicating the presence of plane Rayleigh waves in this frequency band. At 20\,Hz and higher, the mismatch is considerable. There are various explanations for it. For example, reflection of waves from the recess could lead to bias in the array analysis. Some of the seismometers contributing to the array analysis might experience an excess of seismic noise from nearby sources not seen by the tiltmeter. For sure other explanations can be found. At low frequencies, a mismatch is expected since the array-inferred ground tilt requires an estimate of seismic speeds, which we already know to be inaccurate from the dispersion results.

\section{Virgo Newtonian noise reassessment}
\label{Virgo_NN_reassessment}

The array analysis discussed so far is useful to understand the seismic field in terms of its sources, anisotropy, type of waves and their speeds. Typically, these properties lead to mild corrections (by factors $<2$) of NN estimates, but at Virgo, the situation is different. With the presence of clean rooms under the test masses, NN depends strongly on propagation directions \cite{VNNReasses2020}, and above all on seismic speed since the ratio of recess dimensions to seismic wavelengths determines the suppression of NN. For this reason, recesses are ineffective in environments where seismic waves are fast (e.g., underground), and generally, they are more effective at higher frequencies.

Newtonian noise is proportional to the density of the ground medium in the vicinity of a test mass. Hence, building a recess-like structure can reduce the effective density of the environment and thereby reduce NN. Previously, Harms and Hild estimated the possible reduction of NN and also investigated the parameters for the recess structure to maximize the noise reduction. Although the construction of a recess at an existing detector site would be challenging to manage, for sure the construction of recesses should be considered for future surface detectors at new sites \cite{Hall2021} \FLAG{(update reference once published in PRD)}. 

With clean rooms under test masses at the Virgo site, we should expect the gravitational coupling between seismic fields and test masses to be lower than for example at the LIGO sites. In a previous study, we estimated the NN suppression considering a frequency-independent speed of Rayleigh waves of 250\,m/s, which seemed a reasonable choice at that time since similar values had been observed at the two LIGO sites. From the array analysis presented in this paper, we observe that the speed of Rayleigh waves is much lower compared to our previous assumption. In this section, we present results of our analysis of the NN reduction from the recess structure assuming the estimated seismic speed shown in figure \ref{fig:VelsDisp}. From the dispersion curve, we can see that the speed above 10\,Hz lies below 100\,m/s. With such low speed, one should expect more reduction due to the presence of a recess because of the shorter seismic wavelengths.
\begin{figure}[ht!]
  \includegraphics[width=0.95\columnwidth]{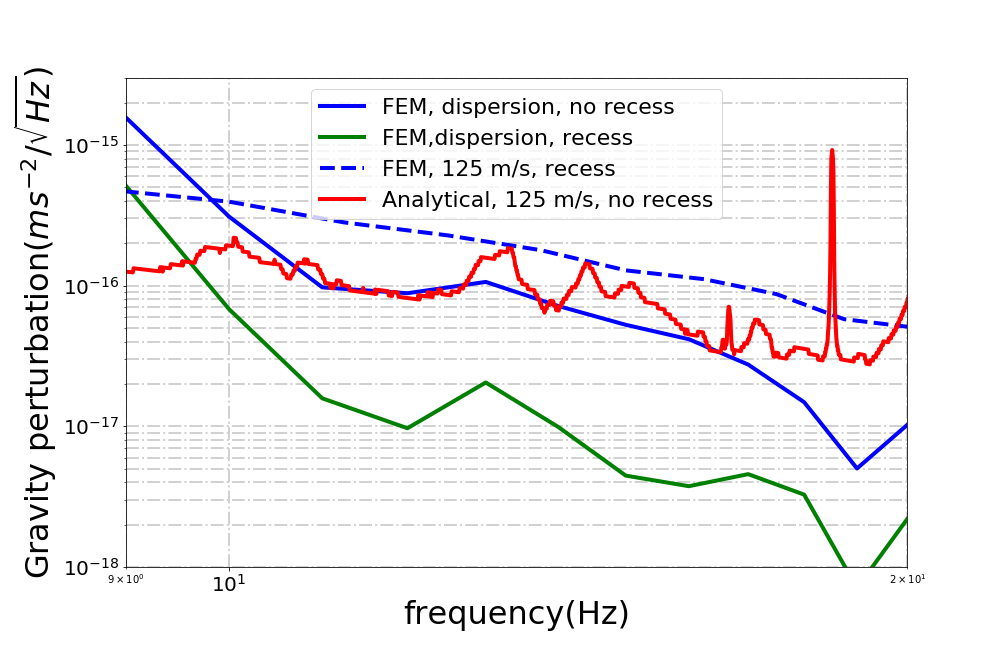}
  \caption{Comparison of NN predictions. The solid blue and green spectra are based on the finite-element simulation without and with recess, respectively, and using the observed Rayleigh-wave dispersion. The dashed spectrum is the result of a finite-element simulation with recess and constant speed. Finally, the red spectrum is an analytical estimate without recess for comparison.}
  \label{fig:GravityPerturbETM_NETM}
\end{figure}
\begin{figure}[bp!] 
    \centering
    \includegraphics[width=0.95\columnwidth]{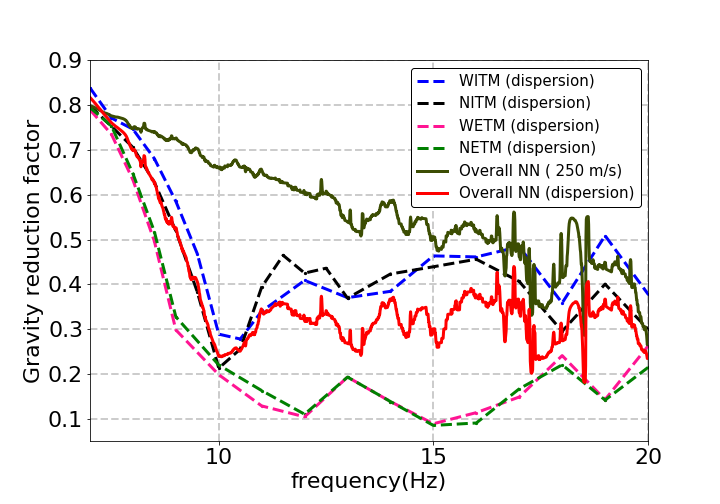}
  \caption{Newtonian-noise reduction from a previous study with a seismic speed of 250\,m/s (olive green) compared to the new results using the measured dispersion curve (red). Reduction factors for each test mass are shown as dashed curves. Here, ITM, ETM mean input and end test mass. The letters W, N stand for West and North arm.}
  \label{fig:ReductionFactors}
\end{figure}
For the first set of results, the assumption is that the seismic field is isotropic. Simulations are carried out with and without recess, and either using the observed wave dispersion or assuming a frequency-independent speed of 125\,m/s. The finite-element simulation is done by propagating plane Rayleigh waves through the model, which means, as explained in detail in our previous study \cite{VNNReasses2020}, that a possible effect of waves being reflected from the clean room is neglected. This might lead to significant errors in our NN estimates above 15\,Hz. The results of the simulations are summarized in figure \ref{fig:GravityPerturbETM_NETM}. We see that the green curve, which is the most accurate prediction using the finite-element simulation with recess and the observed wave dispersion, lies up to a factor 10 below the blue curve, which is the result of an analogous simulation without recess. We also point out that the analytical estimate for a flat surface deviates significantly from the dashed curve at lower frequencies, which is explained by the finite size of the model. This effect is worse when waves are longer. The NN estimate based on the slow observed Rayleigh-wave speed is accurate down to 10\,Hz.

In figure \ref{fig:ReductionFactors}, we show noise-reduction factors for input and end-test masses. Since the seismic speed was measured only at the NEB, the noise reduction at the other two buildings is obtained assuming that the dispersion curve is the same everywhere. We compare with a frequency-independent speed of 250\,m/s, which was the value chosen in our previous study \cite{VNNReasses2020}. It makes it clear how much results have changed between the two studies (olive green curve representing the previous result of NN reduction). Reduction factors are different for input test masses (ITM) and end test masses (ETM) since the recess has a different geometry at Virgo's central building. Reduction is larger at the end buildings.
\begin{figure}[htp!]
    \centering
    \includegraphics[width=0.95\columnwidth]{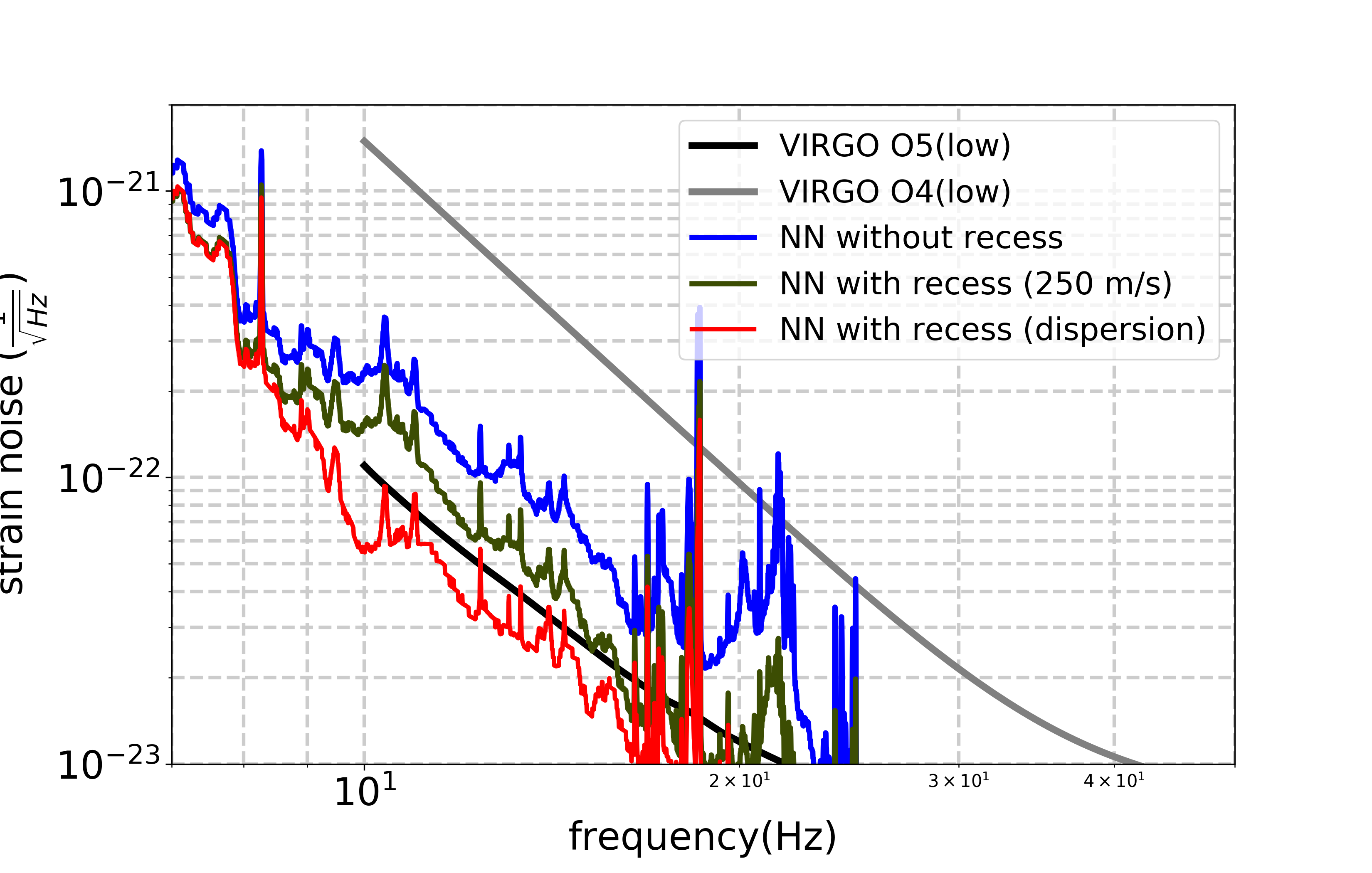}
  \caption{Comparing of Virgo NN estimates summing contributions from all four test masses. For reference, we have also included the Virgo sensitivity models for the two upcoming observation runs O4 and O5 \cite{Abbott2020x}. The blue curve represents the NN for a flat surface. Estimates including the recess are shown for a constant velocity of 250\,m/s (green) and including the estimated Rayleigh-wave dispersion (red).}
  \label{fig:NN_overall}
\end{figure}

As expected, we achieve a greater reduction compared to the previous estimation when incorporating the seismic dispersion.  We also show the NN spectra in figure \ref{fig:NN_overall}. The blue spectrum was computed for a flat surface and shown for comparison to highlight the effect of the recesses inside Virgo buildings. Except for a few peaks in the spectrum, we predict that seismic NN will lie below sensitivity targets of the next two observation runs O4 and O5, but only by a modest factor during O5. This would have important consequences for NNC, which might have to provide only a minor noise reduction according to these results. Certainly, speed measurements at Virgo's West End Building and at its Central Building will be required to corroborate these findings.

\section{Summary and Outlook}\label{Conclusion}
In this paper, we presented an analysis of seismic array data from Virgo's North End Building (NEB). Based on spatial spectra of the field, we measured the dispersion of Rayleigh waves, which we used to update previous Newtonian-noise (NN) predictions for Virgo. We predict NN from seismic fields to lie below the sensitivity targets of the next two observing runs with the exception of a few peaks in the NN spectra. This means that the clean rooms under the Virgo test masses lead to a NN reduction by up to a factor 10, which is an enormous benefit of the Virgo architecture. The suppression is facilitated by the extreme slowness of Rayleigh waves with speeds below 100\,m/s in the NN band. 

We also estimated ground tilt from array data and compared it with the tilt measured by a tiltmeter located at the NEB. We observed a significant mismatch between the two, which indicates that the seismic field is not well represented by a plane Rayleigh-wave model, which is a condition for tiltmeters to perform effective NN cancellation. The presence of the clean rooms might lead to significant scattering of seismic waves above 15\,Hz, which makes a tiltmeter less effective. 

Our results indicate that the construction of recesses under and around test masses can be an efficient way to reduce NN. Constructing recesses as upgrade of existing detectors is challenging since the operation is invasive and requires modification of how vacuum chambers and pipes are mounted to the ground. At new detector sites, it is an attractive technique, but an evaluation of seismic speeds needs to be done in advance to predict the effectiveness of recesses for NN suppression. Recesses are less effective at lower frequencies and higher seismic speeds. This basically renders such methods useless at the proposed underground infrastructure Einstein Telescope. 

\section{Acknowledgement}
TB and BI acknowledge support from grants FNP: TEAM/2016-3/19 and ''AstroCeNT: Particle Astrophysics Science and Technology Centre'' (MAB/2018/7) carried out within the International Research Agendas programme of the Foundation for Polish Science (FNP) financed by the European Union under the European Regional Development Fund.We also want to acknowledge the LIGO and VIRGO clusters, where the analysis was run.

\bibliographystyle{apsrev} 
\bibliography{references}

\end{document}